\documentclass[%
reprint,
superscriptaddress,
 amsmath,amssymb,
 aps, prl
]{revtex4-1}

\usepackage{graphicx}
\usepackage{ulem}

\begin{document}

\title{Observation of Quantum Interferences via Light Induced Conical Intersections in Diatomic Molecules}

\author{Adi Natan}
\affiliation{Stanford PULSE Institute, SLAC National Accelerator Laboratory \\ 2575 Sand Hill Rd. Menlo Park, CA, 94025, USA}
\email{natan@stanford.edu}

\author{Matthew R Ware}
\affiliation{Stanford PULSE Institute, SLAC National Accelerator Laboratory \\ 2575 Sand Hill Rd. Menlo Park, CA, 94025, USA}
\affiliation{Department of Applied Physics, Stanford University, Stanford, California 94305, USA}

\author{Vaibhav S. Prabhudesai}
\affiliation{Department of Physics of Complex Systems, Weizmann Institute of Science,
Rehovot 76100, Israel}

\author{Uri Lev}
\affiliation{Department of Particle Physics and Astrophysics, Weizmann Institute of
Science, Rehovot 76100, Israel}

\author{Barry D. Bruner}
\affiliation{Department of Physics of Complex Systems, Weizmann Institute of Science,
Rehovot 76100, Israel}

\author{Oded Heber}
\affiliation{Department of Particle Physics and Astrophysics, Weizmann Institute of
Science, Rehovot 76100, Israel}



\author{Philip H Bucksbaum}
\affiliation{Stanford PULSE Institute, SLAC National Accelerator Laboratory \\ 2575 Sand Hill Rd. Menlo Park, CA, 94025, USA}
\affiliation{Department of Applied Physics, Stanford University, Stanford, California 94305, USA}
\affiliation{Department of Physics, Stanford University, Stanford, California 94305, USA}

\begin{abstract}
 We observe energy-dependent angle-resolved diffraction patterns in protons from strong-field dissociation of the molecular hydrogen ion H$_2^+$.  The interference is a characteristic of dissociation around a laser-induced conical intersection (LICI), which is a point of contact between two surfaces in the dressed 2-dimensional Born-Oppenheimer potential energy landscape of a diatomic molecule in a strong laser field. The interference magnitude and angular period depend strongly on the energy difference between the initial state and the LICI, consistent with coherent diffraction around a cone-shaped potential barrier whose width and thickness depend on the relative energy of the initial state and the cone apex.  These findings are supported by numerical solutions of the time-dependent Schr\"{o}dinger equation for similar experimental conditions.
\end{abstract}

\pacs{32.80.Xx, 33.80.-b, 82.50.-m}

\maketitle

The Born-Oppenheimer approximation (BOA) represents intramolecular dynamics as the motion of nuclear wave packets on potential energy surfaces (PES) of electronic eigenvalues embedded in the space of nuclear geometries.  A molecule on a single PES remains there so long as the adiabatic condition is obeyed, i.e. so long as nuclear kinetic energies are small compared to electronic state separations. This assumption must break down, however, if two or more PESs approach each other \cite{Baer,Levine,Worth}. When this happens non-adiabatic couplings between the nearly-degenerate surfaces become important.  The true eigenvalues can then be calculated by diagonalizing the Hamiltonian in the reduced space of the near degeneracy.

According to simple geometrical arguments, molecules with at least two dimensions of internal nuclear motion (i.e. three or more atoms) \textit{must} have some points where two or more PESs become degenerate, a condition known as a conical intersection (CI) \cite{Herzberg}. These CIs play a key role in the relaxation dynamics of most polyatomic molecules including important biochemical processes such as the photostability of DNA \cite{McFarland2014}, and the preliminary process of vision \cite{Polli}.  The dimension of the CI manifold is two less than the internal nuclear geometry.  Thus for the simplest case of a tri-atomic molecule, the CI manifold has dimensionality of  $3N-8 = 1$.  The topological nature of CIs allows the nonadiabatic couplings to diverge and thus display related phenomena such as a geometric or Berry's phase \cite{Herzberg,Berry} in wavepackets that circumnavigate the CI.

\begin{figure}
 \includegraphics[width=1.0\columnwidth]{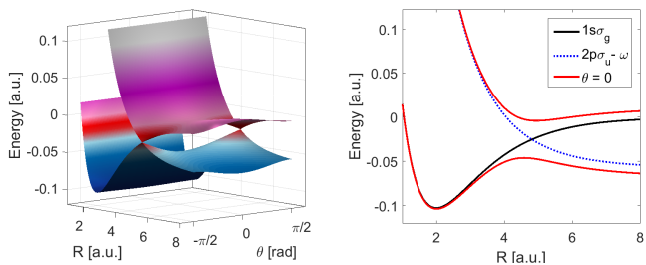}
 \caption{\small PESs of H$_2^+$ and and LICI.  (a) The dressed adiabatic surfaces vs. the interatomic distance R and
angle $\theta$ between the laser polarization and the molecular axis exhibiting the LICI (field intensity: $4 \times 10^{13}$ W/cm$^2$ ).
(b) The diabatic energies of the lower $1s \sigma_g$ (solid black line) and field dressed excited  $2p \sigma_u$ states (dashed blue
line) states of H$_2^+$ form a LICI at R  $\simeq$ 4.8 a.u.  The red curves show a cut at $\theta = 0 $ where adiabatic dynamics prevail.}\label{fig1}
\end{figure}

Naturally occurring CIs cannot exist for a free diatomic molecule because the internuclear separation vector $\mathbf{R}$ is the only internal nuclear degree of freedom, and this is insufficient to fulfil the crossing condition. The non-adiabatic terms in the full Hamiltonian cause the BOA states to repel according to the so-called "no-crossing" rule. A diatomic molecule in a strong laser field, however, has a second degree of freedom defined by the laser polarization $\mathbf{\varepsilon}$. When viewed in a Floquet basis of laser-dressed electronic states,  a molecule coupled by this field can exhibit a point of degeneracy called a light-induced conical intersection (LICI) in the $2-$dimensional space spanned by $[\mathbf{R}, \mathbf{\varepsilon}]$.  The angle of polarization $\theta$ of the laser with respect to the molecular axis spans the missing second degree of freedom that is needed for the crossing condition to take place, as seen in Figure \ref{fig1}. The laser-induced dipole coupling $D$ of the two states also generally leads to avoided crossings in the adiabatic dressed PES landscape; but the gap must vanish at the value of $R$ where the two states come into resonance and at the angle where the dipole moment vanishes. This is the location of the LICI.  A higher laser intensity produces a steeper LICI cone, reflecting a more abrupt transition from adiabatic to non-adiabatic dynamics in the vicinity of the LICI singularity.

The underlying framework of light induced potential dynamics in H$_2^+$ were described in early calculations and observations of strong-field dissociation \cite{Zavriyev1990}. The first description of LICIs for diatomic molecules concerned the molecular interactions with standing laser waves \cite{Moiseyev}, where a periodic array of conical intersections is produced in the phase space of rovibrational and translational motions. Soon after, LICIs were discussed for diatomic molecules interacting with laser pulses \cite{Sindelka}, where the dressed electronic states exhibit a degeneracy in the $[R, \theta ]$ plane.

Recent theoretical studies show that LICIs contribute to non-adiabatic effects between the electronic, vibrational, and rotational degrees of freedom of diatomic molecules exposed to resonant laser pulses in the optical \cite{Moiseyev2011,Halász1,Halász2,Halász3} and  x-ray \cite{Demekhin1,Demekhin2} regimes. There is a growing interest in these concepts applied to trapping and dynamics of ultracold atoms \cite{Rost1}.

A signature of LICI's is quantum interference in strong-field laser-induced dissociation of the hydrogen or deuterium molecular ion \cite{Halász4,Halász5}. Calculations predict that the interference produces modulations of the angular distributions of the dissociated fragments. Here we report energy- and angle-resolved measurements of modulations in angular distributions of strong field photodissociation of  H$_2^+$ and compare our results to the predictions within the LICI framework.


We have examined the photodissociation of H$_2^+$  obtained by focusing a transform-limited 30 fs pulse at 795 nm with a peak intensity of $2\times10^{13}$ W/cm$^2$ into a well-collimated 4 keV   H$_2^+$  beam. The molecular ions are extracted from a Nielsen source \cite{Nielsen} by ionizing H$_2$, and contain a distribution of vibrational states that is close to the predictions of the Franck-Condon approximation. Both H and H$^+$ fragments are measured in coincidence using a time- and position-sensitive detector comprised of a phosphor screen anode attached to a microchannel plate with a CCD camera. The unscattered ion beam is collected in a Faraday cup located in front of a multi-channel plate assembly. The full 3D momentum of the dissociated fragments can be reconstructed from the measured position and time of flight. From these momentum components, we calculate the kinetic energy released and the angle $\theta$ between the molecular axis and the laser polarization at the time of dissociation for each dissociation event. Additional details on the experimental setup can be found elsewhere \cite{Prabhudesai2010,Natan2012}.

\begin{figure}
 \includegraphics[width=1.0\columnwidth]{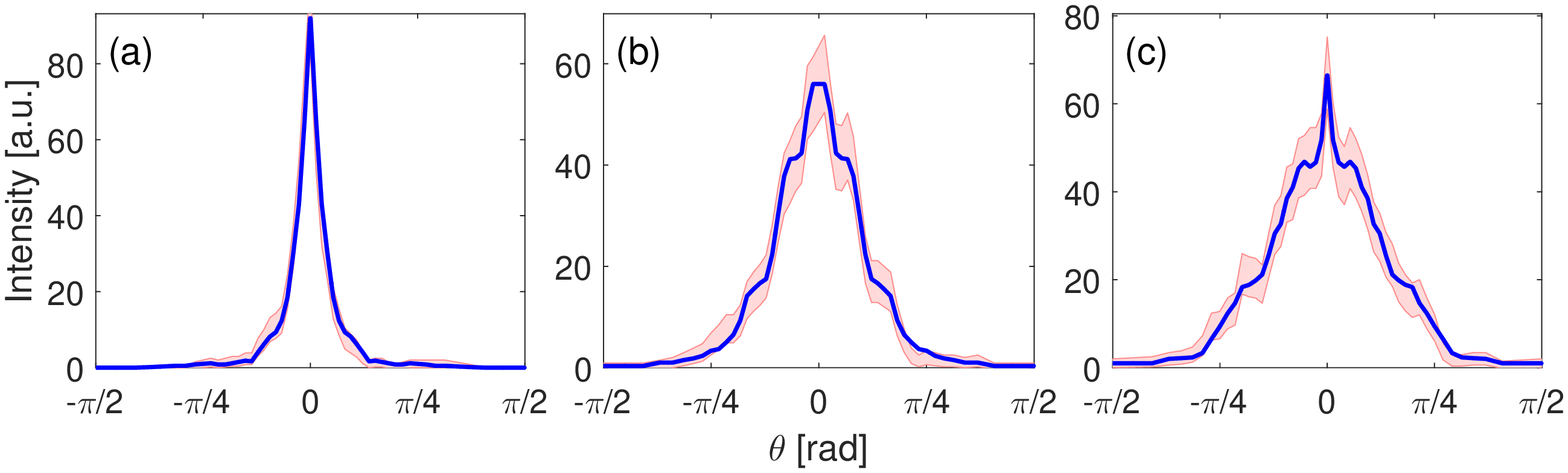}
 \caption{Measured dissociation yield vs angle of H$_2^+$ for different kinetic energy release values that correspond to the vibration levels (a) v=7 at 0.5 eV (b) v=8 at 0.67 eV and (c) v=9 at 0.83 eV. The angular distributions reveal modulations at different angles that increase from v=7 to v=9. The shaded area represent the standard error of the measurement. The solid traces are the angular distributions following symmetrization around $\theta=0$.}\label{fig2prl}
\end{figure}

 We concentrate here on angular distributions at specific kinetic energy releases (KER) from rovibrational states in the vicinity of the LICI.  Fig \ref{fig2prl} contains the measured angular distributions taken from very narrow (25 meV) KER slices that corresponds to the expected average release energy for dissociation from vibration levels $v=7,8,9$ at the ground rotational state. Higher vibration states were not considered in the analysis because of lower overall signal strengths associated with their Franck-Condon factors, and the challange to distinguish their contribution from lower vibration levels with higher rotational quantum number, which have a similar kinetic energy release. A notable feature in Fig \ref{fig2prl} is the increase in the overall width of the angular distributions over this energy release range. This can be understood as follows:  The outer turning point of $v=9$ is at a location on the PES that is close to resonance with the field, and as a result dissociation from this level is mostly induced by resonant one-photon transitions that produce a wide $cos^2(\theta)$ distribution of atomic fragments. In contrast, $v=7$ is not resonant with the laser field, so dissociation occurs via bond-softening. This leads to a much narrower distribution of fragment dissociation angles.  The best fit is dominated by a $cos^8(\theta)$ dependence.

We also observe distinct modulations in the angular distributions, at angles 10$^\circ$-70$^\circ$,  that stand in contrast to the typical smooth $cos^{2n}(\theta)$ type distributions that were previously measured by KER-integrated strong field dissociation \cite{Natan2012,BenItzhak2005}. These modulations also increase in their number and complexity for increasing KER. The modulations are washed out if the KER distribution bin size is increased above 25 meV. The choice of this bin size was to minimize contributions of adjacent rovibration states into the angular distribution that we inspect. So even though the vibrational structure is broad in energy for such laser pulses, the rotational structure that is obtained via the angular distributions can rapidly change as function of energy, as seen in previous studies  \cite{Prabhudesai2010,Natan2012,BenItzhak2005}.

These modulations are consistent with quantum interferences caused by the existence of the LICI. To explore this further we have numerically solved the time-dependent Schr\"{o}dinger equation both for nuclear and rotational degrees of freedom of our molecular system H$_2^+$ using a standard pseudo-spectral method \cite{Fornberg}. In H$_2^+$ only the $1s \sigma_g$ and $2p \sigma_u$ potential curves need to be considered for the field intensities used here.

\begin{figure}
\includegraphics[width=1.0\columnwidth]{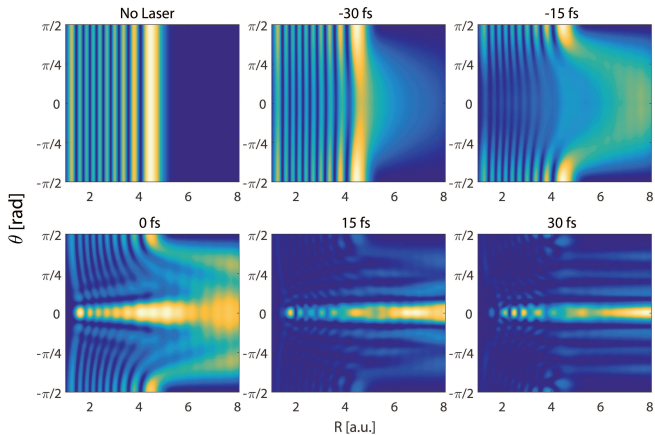} 
\caption{Snapshots of the total probability density as function of internuclear separation $R$ and angle $\theta$. The initial wavefunction was the $|v=9,J=0\rangle$ rovibrational bound eigenstate of H$_2^+$, which depends only on $R$.  As the laser pulse interacts with the molecule, we see a transition from a smooth angular distribution to strong modulations. The position of the LICI is at $\theta= \pm \pi/2 $ and $R\simeq 4.8$. The laser pulse was $30$ fs (FWHM) around $795$ nm and its peak intensity happens at $t=0$ fs.} \label{fig3prl}
\end{figure}

In order to visualize the field-induced dynamics in the vicinity of the LICI, as well as to infer experimentally observable quantities such as the angular distribution of the dissociated system, we compute the probability density as function of the internuclear distance $R$ and the laser polarization angle $\theta$. The initial state of the molecule is a pure rovibrational eigenstate $|v,J,M\rangle$ where $J=0$, which then evolves under the influence of a laser pulse similar to the one used in the experiment. Figure \ref{fig3prl} shows several snapshots of the probability density from -30 fs before the peak of the pulse to $+30$ fs for the initial state $v=9,J=0$.  The numerical analysis demonstrates that at early times ($< -15$  fs) the outgoing part of the probability density (internuclear separation of $R>7$ a.u.) has a $\cos^2(\theta)$ distribution, as expected for resonant dipole coupling to a dissociating curve. Just before the peak of the pulse the probability density in the vicinity of the LICI ( $\theta= \pm \pi/2 $ and $R\simeq 4.8$) is bigger than that around $\theta=0$. The probability density close to the LICI cannot freely dissociate because most of the population is still in the ground state and the LICI acts as a local maximum in the bottom PES, allowing the wavefunction to accumulate near it. This effect couples the rovibrational states, leading to interference structures and narrowing the angular distribution.

The angular structure at high intensity is due to interference between the part of the wavepacket that follows the adiabatic portion of the light-molecule potential around $\theta=0$ and the portion that undergos non-perturbative and non-adiabatic population transfer to higher rovibration states. Phase shifts develop between these different components due to rotational dispersion. Longer effective times of interaction with the LICI lead to larger phase shifts and more developed interference patterns. In Fig \ref{fig3prl}  this is seen at first as a diagonal nodal structure in the $[R,\theta]$ phase space, which evolves into an interference pattern in the angular distribution of the outgoing fragments.

We compute the angular distribution of the dissociating fragments by placing an absorbing boundary at a large internuclear separation to capture the unbound parts of the wavepacket. We calculated the instantaneous dissociation probability by projecting the part of the dissociating wavefunction onto the absorbing boundary along the spatial axis using:

 \[  P(\theta,t)=\int_0^\infty |\Psi(R,\theta,t)|^2 W(R) dR  \]

\begin{figure}
 \includegraphics[width=1.0\columnwidth]{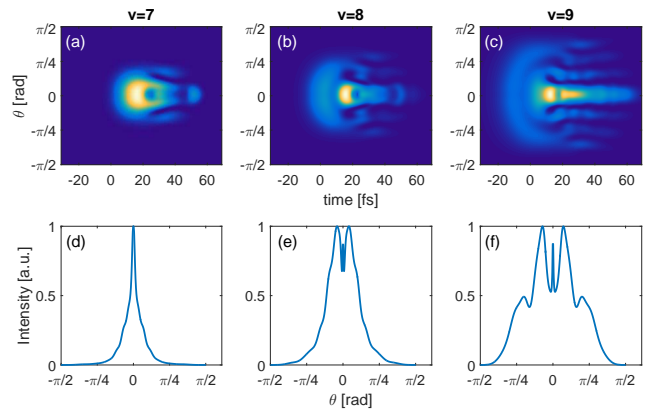}
 \caption{(top) Calculated dissociation probability as function of time and angle for molecules that were initially in the ground rotational state of the   $v=7,8,9$ vibrational states. We note that the closer the initial state is to resonance with the laser field, the more time it interacts with the LICI, resulting in more complex interferences in the angular distributions of the instantaneous dissociation probability. (bottom) Calculated kinetic energy release curves for the same values and energy bin sizes that are presented in Fig 2. Kinetic energy release distributions  were calculated from the dissociating part of the wavepacket of each initial vibration state, and were then normalized by their Franck-Condon coefficient. The interference peaks at $\pm 10^\circ$ and $\pm 35^\circ$ are accurately captured by the model.  Differences in intensity, especially around $\theta=0$ are due to volume averaging effects and finite rotational temperature that are excluded from this calculation.}\label{fig4prl}
\end{figure}

\looseness=-1
 where $W(R)$ is the absorbing boundary function that inhibit back reflections of the propagating wave packet. In  Fig \ref{fig4prl}(a-c), we show the calculated $P(\theta,t)$  for an H$_2^+$ molecule that was initially in the ground rotational state of the   $v=7,8,9$ vibration energies that were accessible in the experiment. Each energy has a different angular distribution and also a different effective interaction time. This result is another demonstration that the high end of the energy range, corresponding to $v=9$ (Fig \ref{fig4prl}(c)), is closest to resonance, while the low energy range, which corresponds to $v=7$ (Fig \ref{fig4prl}(a)) is non-resonant and has a narrower angular distribution. We see that the initial states that are closer to resonance have longer  effective interaction times and are more affected by the existence of the LICI, demonstrated by the number of nodes that develop as function of time in Fig \ref{fig4prl}(a-c) for each vibration level. For example, the initial overlap of $v=7$  with the LICI is negligible.  Significant rotational scattering from the LICI only occurs at the peak of the pulse, where bond softening and non-resonant Raman transitions can take place. As seen in Fig \ref{fig4prl}(a), only a single node evolves at $t=+30$ fs in the range $0<\theta<\pi/2$. The picture changes for $v=8$, where two nodes are noticeable in the same angle range, and the appearance of these nodes start earlier at $t=10$ fs. Finally, for $v=9$, a more developed interference pattern begins to evolve at $t=-10$ fs.

  In order to compare the outcome of the calculation with the experiential results we need to extract the relevant KER distributions from the dissociating part of the wavepackets of each initial vibration state, and weigh them by their Franck-Condon coefficients. Fig 4(d-f) show the calculated KER at the same energy positions and bin sizes as in Fig \ref{fig2prl}, where we found that using a pulse intensity of $3 \times 10^{13}$ W/cm$^2$ gave better agreement with our experiential results . Analyzing the contributions of each initial vibration state on the KER distribution show that different parts of the interferences presented in Fig \ref{fig4prl}(a-c) may end up with a different KER values. This is expected as the bandwidth of the laser pulse used will result in a broad KER structure. Consequently, we can trace the origins of the interference peaks in Fig \ref{fig4prl}(d-f) to the interference structure shown in each of the dissociation probability shown in Fig \ref{fig4prl}(a-c). We establish that the peaks at approximately $\pm 10^\circ$ for  $v=8$ and  $v=9$ are indeed related to interferences seen in the total $P(\theta,t)$. However, the structure around $\pm35^\circ$  for KER=0.83 eV (Fig 4f) does not originate from the $v=9$ state but rather from an interference of the $v=8$ initial state that ended up contributing at that same energy. We therefor expect that the structure at $\pm 60^\circ$ (0.67 eV Fig \ref{fig2prl}(b)) originates from rovibration states of $v=7$ that were not included in the calculation. Interestingly, the strong peak at $\theta=0$ seen at Fig 4(b-c) is spread on a much broader KER range and contribute less when we only consider the $v=7,8,9$, $J=0$ states. In the experiment though, we have other channels that contribute to the detected central peak. For example,  larger iso-intensity shells that are needed to be accounted for focal volume averaging will mainly contribute signals around $\theta=0$  that are not related to the LICI mechanism.  Collectively,  these findings are consistent with the evolution of the phase shifts that were discussed earlier, and also agree with the observation of the modulations at the three energies shown in Fig \ref{fig2prl}.  We therefor identify these modulations as the quantum interferences that are caused by the existence of the LICI.

 Though there is an agrement between the simulation and the experimental results with respect to the positions of the interferences, there are quantitative differences due to a number of technical limitations and non-ideal experimental condition. For example, the simulation assumes a cold rotational sample ($J=0$), whereas, in this experiment the molecular ions created were hot with an undetermined rotational temperature and only an approximate estimation of its vibrational occupation. One consequence is that the ensemble of molecules had a finite population of $|J,M\rangle$ states that add incoherently in all angles.  However, for a hot rotational ensemble, there will always be additional contributions in the same final kinetic energy from other rovibration states. In order to minimize this effect we have looked at a thin KER region where the expected contribution from the $J=0$ is largest. In the calculation we have also neglected focal volume averaging. This will change angular distributions by adding more adiabatic type dissociation events, decreasing the contrast of the interferences and adding to the distribution signal at smaller angles. Nevertheless,  we note that similar quantum interferences in photodissociation of diatomic molecules were also theoretically discussed very recently in the framework of LICI \cite{Halász4,Halász5} and our analysis is consistent with these studies.

In summary, we have experimentally demonstrated the effects of light-induced conical intersections (LICIs) on strong-field photo- dissociation of H$_2^+$ by means of quantum interferences that modulate the angular distributions of the fragments. The effect depends on the effective  interaction time between the LICI and the dissociating wave packet.  Initial states that are closer to resonance are more strongly affected and develop richer interference structures. These observations are supported by a calculation of the system under similar conditions.  The non-adiabatic effects emerging from LICIs should be amenable to control because of possibility to shape the properties of the LICI by laser frequency, duration, intensity, and polarization. The concept of LICIs can  also be applied to polyatomic molecules, as shown in a recent laser-induced isomerization experiment \cite{Kim2012}. The main difference here compared to the case of diatomic molecules is that several internal nuclear degrees of freedom may be relevant for the formation of the  LICI.  This is particularly  advantageous  when considering that naturally occurring CIs are more difficult to control, because their positions and characteristics are inherent to the specific molecular system.

This research is supported through Stanford PULSE Institute, SLAC National Accelerator Lab by the U.S. Department of Energy, Office of Basic Energy Sciences, Atomic, Molecular, and Optical Science Program.


\end{document}